\documentclass[aps,prd,preprint,nofootinbib,preprintnumbers,superscriptaddress,floatfix]{revtex4}

\usepackage{epsfig}
\usepackage{bm}
\usepackage{amssymb}
\usepackage{amsmath}
\usepackage{color}
\usepackage{subfigure}
\usepackage{dcolumn}

\usepackage[utf8]{inputenc}

\usepackage{amsmath}
\usepackage{amssymb}
\usepackage{amsthm}
\usepackage{amscd, amsfonts, mathrsfs}
\usepackage{cases}
\usepackage{verbatim} 
\usepackage{epsf}
\usepackage[bookmarksnumbered,pdfpagelabels=true,plainpages=false,colorlinks=true,
            linkcolor=black,citecolor=black,urlcolor=black]{hyperref}

\theoremstyle{plain}

\theoremstyle{definition}

\allowdisplaybreaks

\newcommand{\cQ}{\mathcal Q}

\newcommand{\Bracket}[1]{\left \langle #1 \right \rangle }

\newcommand{\MA}{\mathcal{A}}

\newcommand{\bml}{\begin{subequations}}
\newcommand{\eml}{\end{subequations}}

\newcommand{\be}{\begin{equation}}
\newcommand{\ee}{\end{equation}}
\newcommand{\bea}{\begin{eqnarray}}
\newcommand{\eea}{\end{eqnarray}}

\newcommand{\bbm}{\begin{bmatrix}}
\newcommand{\ebm}{\end{bmatrix}}
\newcommand{\bvm}{\begin{vmatrix}}
\newcommand{\evm}{\end{vmatrix}}

\begin{document}

%%%%%%%%%%%%%%%%%

\title{Nonlinear causality of general first-order relativistic viscous hydrodynamics}
\date{\today}

\author{F\'abio S.\ Bemfica}
\affiliation{Escola de Ciências e Tecnologia, Universidade Federal do Rio Grande do Norte, 59072-970, Natal, RN, Brazil}
\email{fabio.bemfica@ect.ufrn.br}

\author{Marcelo M.\ Disconzi}
\affiliation{Department of Mathematics, Vanderbilt University, Nashville, TN, USA}
\email{marcelo.disconzi@vanderbilt.edu}

\author{Jorge Noronha}
\affiliation{Department of Physics, University of Illinois, 1110 W. Green St., Urbana IL 61801-3080, USA}
\email{jn0508@illinois.edu}
\affiliation{Instituto de Física, Universidade de São Paulo, Rua do Matão, 1371, Butantã, CEP 05508-090, São Paulo, SP, Brazil}
%\date{\today}
%\pagestyle{empty}

\begin{abstract}
Effective theory arguments are used to derive the most general energy-momentum tensor of a relativistic viscous fluid with an arbitrary equation of state (in the absence of other conserved currents) that is first-order in the derivatives of the energy density and flow velocity and does not include extended variables such as in Mueller-Israel-Stewart-like theories. This energy-momentum tensor leads to a causal theory, provided one abandons the usual conventions for the out-of-equilibrium hydrodynamic variables put forward by Landau-Lifshitz and Eckart. In particular, causality requires nonzero out-of-equilibrium energy density corrections and heat flow. Conditions are found to ensure linear stability around equilibrium in flat space-time. We also prove local existence and uniqueness of solutions to the equations of motion. Our causality, existence, and uniqueness results hold in the full nonlinear regime, without symmetry assumptions, in four space-time dimensions, with or without coupling to Einstein's equations, and are mathematically rigorously established. Furthermore, a kinetic theory realization of this energy-momentum tensor is also provided. 
\end{abstract}

\maketitle

%%%%%%%%%%%%%%% INTRODUCTION %%%%%%%%%%%%%%%%%%%%%%

\section{Introduction} 

Relativistic fluid dynamics plays an important role in high-energy nuclear physics \cite{Heinz:2013th}, astrophysics \cite{RezzollaZanottiBookRelHydro}, and cosmology \cite{WeinbergCosmology}. Its wide range of applicability stems from the application of general conservation laws in situations where there is a large hierarchy among length scales, so that the macroscopic behavior of conserved quantities (such as energy and momentum \cite{LandauLifshitzFluids}) can be determined without detailed information about the system's underlying microscopic dynamics. 

Ideal hydrodynamic behavior, corresponding to the limit where dissipation can be neglected, is physically well understood \cite{RezzollaZanottiBookRelHydro,LandauLifshitzFluids}. 
In the absence of other conserved currents (i.e., at zero chemical potential), an ideal relativistic fluid can be described using the energy-momentum tensor $T^{\mu\nu}_{ideal} = \varepsilon u^\mu u^\nu + P(\varepsilon)\Delta^{\mu\nu}$, where $\varepsilon$ is the energy density, $P=P(\varepsilon)$ is the equilibrium pressure defined by the thermodynamic equation of state, $u^\mu$ (with $u^\mu u_\mu = -1$) is the local flow velocity, $\Delta_{\mu\nu} = g_{\mu\nu} + u_\mu u_\nu$ is a projector orthogonal to $u^\mu$, and $g_{\mu\nu}$ is the space-time metric. The dynamics of the fluid is determined by solving the relativistic Euler equations defined by energy-momentum conservation, i.e., $\nabla_\mu T^{\mu\nu}_{ideal}=0$, which give first-order equations of motion for the hydrodynamic variables $\{\varepsilon,u^\mu\}$. It is known that the equations of motion are locally well-posed, i.e., given suitable initial data for the variables a 
unique solution exists, and that causality (defined below) also holds \cite{AnileBook}.
In the more general case where gravitational effects cannot be neglected
\cite{RezzollaZanottiBookRelHydro}, the metric is determined by Einstein's equations and the initial value problem for the Einstein-Euler  is also locally well-posed 
and causal \cite{Choquet-BruhatFluidsExistence, DisconziRemarksEinsteinEuler}. 

Saying that causality holds for a system of equations means that 
the values of a solution at a given space-time point $x$ are completely determined by the space-time region that is in the past of and causally connected to $x$ \cite{ChoquetBruhatGRBook,WaldBookGR1984}. In other words, causality implies that information cannot propagate at superluminal speeds. Given that this concept is central in relativity, it must also hold when dissipative phenomena are taken into account. However, relativistic causality and dissipation in fluid dynamics have been at odds since the
work of Eckart \cite{EckartViscous} in 1940.

In this work we investigate the most general expression for the energy-momentum tensor of a relativistic viscous fluid at zero chemical potential, with an arbitrary equation of state, where dissipative corrections are taken into account via first-order derivatives of the energy density and flow velocity. 
Theories where dissipative effects are modeled in this way are traditionally referred to as
first-order theories. We go beyond all previous results concerning relativistic viscous hydrodynamics by proving causality, local existence, and uniqueness of solutions to Einstein's equations coupled to this most general viscous fluid in the nonlinear regime. We show that causality requires nonzero out-of-equilibrium energy density corrections and heat flow. Without these ingredients, our theory reduces to that of Landau and Lifshitz \cite{LandauLifshitzFluids}, which is known to be acausal. Comprehensive conditions are found to ensure linear stability around equilibrium in flat space-time. Furthermore, we show how the general energy-momentum introduced here can be derived from kinetic theory.

This paper is organized as follows. In the next section we briefly review the previous approaches to relativistic viscous fluid dynamics. Section \ref{generalTensor} provides a derivation of the most general viscous energy-momentum tensor at first-order and discusses our proof of causality, local existence, and uniqueness of solutions to the equations that describe the viscous fluid and its coupling to Einstein's equations. A linear stability analysis around hydrostatic equilibrium in Minkowski space-time is also presented in this section. We finish the paper with our conclusions and outlook in Section \ref{conclusions}. Appendix \ref{Appendix_kinetic} shows how the energy-momentum tensor studied here can be derived from kinetic theory while in Appendix \ref{Appendix_causality} we discuss the formal aspects of the proofs and give the necessary technical mathematical details. We use units where $c = \hbar = k_B = 1$. The space-time metric signature is $(-+++)$. Greek indices run from 0 to 3, Latin indices from 1 to 3.

\vskip 0.1cm
\noindent \textbf{Note added:} While we were finishing this paper, 
we became aware of \cite{Kovtun:2019hdm}, which also investigated stability
and causality (in the linear regime) of the energy-momentum tensor in \eqref{novotensorBDNfodaotche} and \eqref{BDNdef}.   
\vskip 0.1cm

%%%%%%%%%%%%%% SECTION  %%%%%%%%%%%%%%%%%%%%%%%%%%%%

\section{Previous approaches}

Formulations of viscous relativistic fluid dynamics were first proposed by Eckart \cite{EckartViscous} and Landau and Lifshitz \cite{LandauLifshitzFluids}. Given that $u_\mu T^{\mu\nu}_{ideal} = -\varepsilon u^\nu$, the Landau-Lifshitz theory assumes that the same relation holds when dissipation is included. The most general energy-momentum tensor for a fluid that satisfies this condition is $T^{\mu\nu} = \varepsilon u^\mu u^\nu + (P+\Pi)\Delta^{\mu\nu} + \pi^{\mu\nu}$, where $\Pi$ is the bulk scalar and $\pi^{\mu\nu}$ is the shear stress tensor, $\pi_{\mu\nu} = \Delta^{\alpha\beta}_{\mu\nu} T_{\alpha\beta}$, where $\Delta^{\alpha\beta}_{\mu\nu} = \left(\Delta^\alpha_\mu \Delta^\beta_\nu+\Delta^\alpha_\nu \Delta^\beta_\mu\right)/2 -  \Delta^{\alpha\beta}\Delta_{\mu\nu}/3$. In equilibrium $\Pi$ and $\pi^{\mu\nu}$ vanish and one returns to ideal hydrodynamics. Assuming that the only degrees of freedom are still the hydrodynamic fields already defined in the ideal case, small deviations from local equilibrium described by $\Pi$ and $\pi^{\mu\nu}$ can be written as an expansion in powers of the space-time derivatives of $\{\varepsilon,u^\mu\}$. This is known as the gradient expansion in fluid dynamics \cite{ChapmanCowling,degroot, kremer,Baier:2007ix}. When truncating this expansion at first order in the Landau-Lifshitz theory, one finds $\Pi = - \zeta \nabla_\mu u^\mu$ and $\pi_{\mu\nu} = -2 \eta \sigma_{\mu\nu}$, where $\sigma_{\mu\nu} = \Delta_{\mu\nu}^{\alpha\beta}\nabla_\alpha u_\beta$. The second law of thermodynamics \cite{LandauLifshitzFluids} then implies that the shear and bulk viscosities, $\eta$ and $\zeta$, respectively, are non-negative. After making this choice for the dissipative fields, energy-momentum conservation then gives equations of motion that provide a possible relativistic generalization of the classical Navier-Stokes equations \cite{LandauLifshitzFluids}. Despite being physically motivated, this theory is acausal \cite{PichonViscous} and unstable \cite{Hiscock_Lindblom_instability_1985}. Such pathologies are very severe, especially in the context of general relativity applications (Eckart's theory has the same problems).
In fact, the results of \cite{Hiscock_Lindblom_instability_1985} hold
for a large (but not exhaustive) class of first-order theories, leading to a widespread belief
that causality and stability could not be accomplished in the framework of first-order theories.

A possible solution to this long-standing acausality problem was proposed by Mueller, Israel, and Stewart (MIS) \cite{MIS-1,MIS-2,MIS-6} decades ago. Again, the energy-momentum tensor of the viscous fluid at zero chemical potential is assumed to obey the Landau-Lifshitz condition $u_\mu T^{\mu\nu} = -\varepsilon u^\nu$ but now the dissipative fields, $\Pi$ and $\pi^{\mu\nu}$, are found by solving new equations of motion that couple these variables
to the other hydrodynamic fields. The new equations of motion for such new variables
are typically postulated based on some general physical principle such as the 
second law of thermodynamics. A solution to the full set of equations 
of motion requires specifying initial data for the extended set of variables $\{\varepsilon,u^\mu,\Pi, \pi^{\mu\nu}\}$. Theories of this type, based on the developments put forward in \cite{Baier:2007ix} and \cite{Denicol:2012cn}, 
have been successfully used to describe the quark-gluon plasma formed in heavy ion collisions (see \cite{Romatschke:2017ejr} for a review). 

However, it is important to stress that, apart from statements regarding causality (and stability) valid only in the linearized regime \cite{Hiscock_Lindblom_stability_1983, Olson:1989ey,Denicol:2008ha,Pu:2009fj}, it is not known if causality indeed holds under general conditions 
for MIS theories\footnote{Causality has also been studied in the context of the so-called divergence-type theories \cite{GerochLindblomDivergenceType, GerochLindblomCausal,LiuMullerRuggeri-RelThermoGases,MuellerRuggeriBook}. Examples of fluid dynamic theories constructed in this approach can be found in \cite{RamosCalzettaDT1, RamosCalzettaDT2,Lehner:2017yes} (additionally, see \cite{Nagy_et_all-Hyperbolic_parabolic_limit,Kreiss_et_al,Reula_et_al-CausalStatistical}).}. In fact, pathologies associated with nonlinear behavior were observed before in \cite{Hiscock_Lindblom_pathologies_1988}. Moreover, questions regarding the existence and uniqueness of solutions, including the case when the fluid is coupled to gravity, remain open with the exception of highly symmetric situations \cite{Maartens:1995wt}. In this regard, hyperbolicity and causality in MIS theories including shear and bulk viscosities were investigated in \cite{Floerchinger:2017cii} assuming an azimuthally symmetric and boost invariant expansion. To this date, the only general statement regarding causality and well-posedness of solutions in the nonlinear regime in MIS theories, without assuming any simplifying symmetry or near-equilibrium behavior, was recently proven in \cite{BemficaDisconziNoronha_IS_bulk} for the case where only bulk viscosity is included. Therefore, it is not known if the MIS mechanism is powerful enough (or needed) to solve the acausality (and well-posedness) problem of relativistic viscous fluid dynamics under general conditions in the nonlinear regime. In this regard, in order to describe the rapid expansion and the highly anisotropic initial state of the matter formed in heavy ion collisions, a different way to generalize the MIS framework involving a nontrivial resummation of dissipative stresses called anisotropic hydrodynamics \cite{Florkowski:2010cf,Martinez:2010sc} was derived. This approach is rapidly being developed (for a review, see \cite{Alqahtani:2017mhy}) and successful comparisons to heavy-ion data have already been made \cite{Almaalol:2018gjh}. However, precise statements concerning causality and well-posedness in this framework are not known.

%%%%%%%%%%%%%% SECTION  %%%%%%%%%%%%%%%%%%%%%%%%%%%%

\section{General energy-momentum tensor at first-order}\label{generalTensor}

Here, we take a different approach to the problem of acausality in relativistic viscous fluids.
Our approach is motivated by \cite{BemficaDisconziNoronha},
where a first-order stable, causal, and locally well-posed theory was introduced.
However, the work \cite{BemficaDisconziNoronha} was restricted to conformal fluids,
so that it was not clear if causality could indeed be a general feature of first-order theories, as
we show here, or if it was a consequence of the severe constraints imposed by conformal invariance. 

The starting point is that away from equilibrium quantities such as the local temperature $T$ and $u^\mu$ are not uniquely defined \cite{MIS-6} and different choices differ from each other by gradients of the hydrodynamic variables \cite{Kovtun:2012rj}, each particular choice being called a \emph{hydrodynamic frame}. Different frames have been studied over the years by Eckart \cite{EckartViscous}, Landau \cite{LandauLifshitzFluids}, Stewart \cite{Stewart:1972hg} and others \cite{Tsumura:2006hn,Van:2007pw,Tsumura:2011cj,Monnai:2018rgs}. Therefore, a priori, one is not forced to define the hydrodynamic variables such that the Landau-Lifshitz condition $u_\mu T^{\mu\nu} = -\varepsilon u^\nu$ holds out of equilibrium. If this condition is lifted, the most general energy-momentum tensor for the fluid \cite{BemficaDisconziNoronha} is  $T^{\mu\nu}  =\left(\varepsilon+\mathcal{A}_1\right)u^\mu u^\nu +\left(P(\epsilon)+\mathcal{A}_2\right)\Delta^{\mu\nu}+ \pi^{\mu\nu} +\mathcal{Q}^\mu u^\nu + \mathcal{Q}^\nu u^\mu$, where $\mathcal{A}_1$ and $\mathcal{A}_2$ are the non-equilibrium corrections to the energy density and equilibrium pressure, respectively, and $\mathcal{Q}^\mu = -\Delta_\nu^\mu T^{\nu \alpha}u_\alpha$ is the heat flow.

Instead of treating the non-equilibrium corrections as new degrees of freedom (and consequently postulating additional equations for them) as in MIS theories and extended irreversible thermodynamics \cite{JouetallBook}, here we consider the case where the effective theory describing the macroscopic motion of the system is defined solely in terms of $\{\varepsilon,u^\mu\}$. In this case, $\{\mathcal{A}_1,\mathcal{A}_2,\mathcal{Q}^\mu,\pi^{\mu\nu}\}$ must be given in terms of the hydrodynamic fields $\{\varepsilon,u^\mu\}$ and their derivatives, which may be organized through a gradient expansion \cite{Baier:2007ix}. Assuming that deviations from equilibrium are small, the most general theory compatible with the symmetries that can be written at first-order in gradients is given by 
\be
T^{\mu\nu}=(\varepsilon+\MA_1)u^\mu u^\nu+\left (P(\varepsilon)+\MA_2\right )\Delta^{\mu\nu}-2\eta\sigma^{\mu\nu}+u^\mu\mathcal{Q}^\nu+u^\nu\mathcal{Q}^\mu,
\label{novotensorBDNfodaotche}
\ee
where
\be
\MA_1=\chi_1\frac{u^\alpha\nabla_\alpha \varepsilon}{\varepsilon+P}+\chi_2 \nabla_\alpha u^\alpha,\qquad \MA_2=\chi_3\frac{u^\alpha\nabla_\alpha \varepsilon}{\varepsilon+P}+\chi_4 \nabla_\alpha u^\alpha,\qquad 
\mathcal{Q}_\mu=\lambda\left (\frac{c_s^2\Delta^{\nu}_\mu\nabla_\nu \varepsilon}{\varepsilon+P}+u^\alpha\nabla_\alpha u_\mu\right)
\label{BDNdef}
\ee
where $\lambda,\eta,\chi_a$, $a=1,2,3,4$ are transport coefficients 
which are known functions of $\varepsilon$, and 
$c_s^2=d P(\varepsilon)/d\varepsilon$ is the speed of sound squared. We assume that $0\le c^2_s<1$. The coefficients $\lambda,\chi_a$ regularize the ultraviolet behavior of the collective modes of the system in such a way that causality and stability hold. In fact, at the linear level one can show that $\lambda/(\varepsilon+P)$ acts as a type of regulator of high momentum shear modes, playing the same role as the shear relaxation time in MIS theories \cite{Romatschke:2017ejr}. A similar effect occurs in the sound channel, although in a less transparent way. Finally, we note that the conformal tensor proposed in \cite{BemficaDisconziNoronha} is recovered when $P=\varepsilon/3$ and $\chi_1=\chi_2 = \chi$ and $\chi_3=\chi_4=\chi/3$ in Eq.\ \eqref{BDNdef}.

The general expression above fulfills the idea that hydrodynamics can be understood as an effective theory that describes the near-equilibrium behavior of interacting matter at scales where the only relevant degrees of freedom are the standard hydrodynamic fields. As such, this effective theory should be valid for both weakly and strongly coupled systems. Also, we note that since the entropy density is $s = (\varepsilon+P)/T$ \cite{LandauLifshitzFluids},
 we have
$\nabla_\alpha T/T=\nabla_\alpha P/(\varepsilon+P)=c_s^2\nabla_\alpha\varepsilon/(\varepsilon+P)$. Thus, one could also have used $\{T,u^\mu\}$ as variables, as it naturally occurs in kinetic theory \cite{BemficaDisconziNoronha}. We can further use kinetic theory to determine the transport coefficients in \eqref{novotensorBDNfodaotche}-\eqref{BDNdef}. This is shown in Appendix \ref{Appendix_kinetic}, where we derive \eqref{novotensorBDNfodaotche} from kinetic theory. This derivation, in particular, gives that both terms in $\cQ^\mu$ are multiplied by the same transport coefficient (this also follows more generally from the imposition that this term correctly vanishes in thermodynamic equilibrium, as shown in \cite{Kovtun:2019hdm}). Kinetic theory also shows that only three out of the six transport coefficients are independent. 

Additionally, we remark that even though the pressure corrections seem more complicated than the standard $-\zeta \nabla_\mu u^\mu$ expression, the long wavelength behavior of sound disturbances around hydrostatic equilibrium in this theory is given by $\omega_{sound}(k) = \pm c_s k - i\left(\frac{2}{3 T}\frac{\eta}{s} + \frac{1}{2T}\frac{\zeta}{s}\right)k^2 + \mathcal{O}(k^3)$, where $k = \sqrt{k^i k_i}$ and the bulk viscosity is identified as $\zeta = \chi_3 - \chi_4 + c_s^2(\chi_2-\chi_1)$. Shear disturbances are found to be $\omega_{shear}(k) = - i \frac{\eta}{s}\frac{k^2}{T}+\mathcal{O}(k^4)$ and, thus, the long wavelength behavior of this theory near equilibrium is the same as Landau-Lifshitz theory \cite{Romatschke:2017ejr} (we note that the coefficient $\lambda$ only enters at higher orders in the expansion). In fact, as argued in \cite{Kovtun:2019hdm}, in the domain of validity of the equations (i.e., imposing that $T^{\mu\nu}$ is accurate to 1st order) entropy production equals the known expression from Landau-Lifshitz theory \cite{LandauLifshitzFluids} and becomes non-negative if $\eta, \zeta \geq 0$ (there are no further conditions on the other coefficients from this entropy argument). 

However, differently than Landau-Lifshitz theory, the equations of motion obtained from $\nabla_\mu T^{\mu\nu}=0$ with the energy-momentum tensor given by \eqref{novotensorBDNfodaotche} and \eqref{BDNdef} lead to causal propagation, even in the fully nonlinear regime. As a matter of fact, causality only holds when both the heat flow and the non-equilibrium corrections to the energy density (which are both set to zero in Landau-Lifshitz theory) are taken into account. In the next section we present the proof of causality, local existence, and uniqueness of solutions to the equations of motion of this new theory. To motivate further studies of viscous fluid dynamics in the presence of strong gravitational fields in astrophysics and cosmology, the viscous fluid equations are coupled to Einstein's equations. 

%%%%%%%%%%%%%% SECTION  %%%%%%%%%%%%%%%%%%%%%%%%%%%%

\subsection{Causality}\label{section_causality}

In this section we prove that causality holds in the nonlinear regime for the coupled Einstein-viscous fluid system of equations when $\lambda,\chi_1>0$, $\eta \geq 0$,  and conditions \eqref{9} and \eqref{10} below are satisfied, which is the main result of this section. Local existence and uniqueness of the solutions to the equations of motion are also proven below.   

In order to study causality, we need to 
consider the principal part of the system, which is obtained by 
retaining the terms of highest order in derivatives in the equations of motion $\nabla_\nu T^{\mu\nu}=0$ and Einstein's equations  $R_{\mu\nu}-(1/2)g_{\mu\nu}R+\Lambda g_{\mu\nu}=8\pi G\,T_{\mu\nu}$ (where $\Lambda$ is the cosmological constant, added here for completeness) \cite{BemficaDisconziNoronha}. 
In view of the constraint  $u^ \alpha u_\alpha=-1$, only three components
of $u^\mu$ are in fact independent. It is more convenient, however, to treat
all the components $u^\mu$ on the same footing, using the constraint instead 
to split the energy-momentum tensor conservation equation into five equations $u_\mu\nabla_\nu T^{\mu\nu}=0$ and $\Delta^ \alpha_\mu\nabla_\nu T^{\mu\nu}=0$, and we must use the constraint explicitly in the development. Then, the complete set of equations of motion (expressed in wave gauge) can be written as
\bml
\label{4}
\bea
&&\frac{\chi_1 u^\alpha u^\beta+c_s^2\lambda \Delta^{\alpha\beta}}{\varepsilon+P}\partial_\alpha\partial_\beta \varepsilon+\left (\chi_2 +\lambda \right )u^{(\beta}\delta^{\alpha)}_\nu\partial_\alpha\partial_\beta u^\nu+\tilde{B}(\varepsilon,u,g)\partial^2 g=B(\partial \varepsilon,\partial u,\partial g),\label{4a}\\
&&\frac{(\chi_3 +c_s^2\lambda)u^{(\alpha} \Delta^{\mu\beta)} }{(\varepsilon+P)}
\partial_\alpha \partial_\beta \varepsilon +B^{\mu\alpha\beta}_\nu\partial_\alpha
\partial_\beta u^\nu+\tilde{B}^\mu(\varepsilon,u,g)\partial^2 g=B^\mu(\partial \varepsilon,\partial u,\partial g),\label{4b}\\
&&g^{\alpha\beta}\partial_\alpha \partial_\beta g_{\mu\nu}=B_{\mu\nu}(\partial \varepsilon,\partial u,\partial g).\label{4c}
\eea
\eml
where $\tilde{B}(\varepsilon, u, g)\partial^2 g$ and $\tilde{B}^\mu(\varepsilon, u, g)\partial^2 g$ contain all terms of 2nd order in derivatives of the metric $g$ and order zero in $\varepsilon$, $u^\mu$, and $g_{\mu\nu}$, while $B(\partial \varepsilon,\partial u,\partial g)$, $B^\mu(\partial \varepsilon,\partial u,\partial g)$, and $B_{\mu\nu}(\partial \varepsilon,\partial u,\partial g)$ contain all terms with derivatives of order no greater than one (the exact form of $\tilde{B}$ and $B$ will not be relevant for our purposes). Also, we defined 
$B^{\mu\alpha\beta}_\nu=\frac{3\chi_4-\eta}{3}\Delta^{\mu(\beta}\delta^{\alpha)}_\nu+(\lambda u^\alpha u^\beta-\eta  \Delta^{\alpha\beta})\delta^\mu_\nu$.
By constructing the vector $U=(\varepsilon,u^\alpha,g_{\mu\nu})^T\in\mathbb{R}^{15}$ (we consider only the 10 independent $g_{\mu\nu}$), we may write \eqref{4} in matrix form as $\mathcal{M}^{\alpha\beta}\partial^2_{\alpha\beta}U=\mathcal{B}$, where $\mathcal{B}=(B,B_\mu,B_{\mu\nu})\in\mathbb{R}^{15}$ and
\be
\label{5}
\mathcal{M}^{\alpha\beta}=\bbm \mathfrak{m}^{\alpha\beta} & \mathfrak{b}^{\alpha\beta}\\ 0_{10\times 5} & g^{\alpha\beta}I_{10}\ebm
\ee
is a $15\times 15$ real matrix. For simplicity, we define
\be
\label{6}
\mathfrak{m}^{\alpha\beta}=\bbm
\frac{\chi_1 u^\alpha u^\beta+c_s^2\lambda \Delta^{\alpha\beta}}{\varepsilon+P} & \left (\chi_2 +\lambda \right )u^{(\beta}\delta^{\alpha)}_\nu\\
\frac{(\chi_3 +c_s^2\lambda)u^{(\alpha} \Delta^{\mu\beta)} }{(\varepsilon+P)} & B^{\mu\alpha\beta}_\nu
\ebm
\ee
while $\mathfrak{b}$ is a $5\times10$ matrix written in terms of the $\tilde{B}$'s. 

Let $\xi$ be an arbitrary co-vector in space-time. 
To establish causality, we need to verify the following 
\cite{Leray_book_hyperbolic}. For each non-zero $\xi$, the 
roots $\xi_0 = \xi_0(\xi_1,\xi_2,\xi_3)$ of  
$\det(\mathcal{M}^{\alpha\beta}\xi_\alpha\xi_\beta)=0$ are all real
and define a cone, given by the set  $\{ \xi : \xi_0 = \xi_0(\xi_1,\xi_2,\xi_3) \}$,
that lies outside\footnote{\emph{Outside} because $\xi$ is a co-vector, so the
discussion here is in momentum space. By duality, the corresponding 
cone in physical space will be \emph{inside} the light cone.} or equals the lightcone $g^{\alpha\beta} \xi_\alpha \xi_\beta = 0$.

From \eqref{5} it is straightforward to see that the terms in $\mathfrak{b}$ do not contribute and that $\det(\mathcal{M}^{\alpha\beta}\xi_\alpha\xi_\beta)=
(g^{\mu\nu} \xi_\mu \xi_\nu)^{10}\det(\mathfrak{m}^{\alpha\beta}\xi_\alpha\xi_\beta)$. 
The roots coming from the gravity sector, namely, $g^{\alpha\beta} \xi_\alpha \xi_\beta=0$, give the light cones. For the matter sector, we obtain 
$\det(\mathfrak{m}^{\alpha\beta}\xi_\alpha\xi_\beta)=\frac{\lambda^4\chi_1}{(\varepsilon+P)}\prod_{a=1,\pm}\left [(u^ \alpha\xi_\alpha)^2-\tau_a \Delta^{\alpha\beta}\xi_\alpha\xi_\beta\right ]^{n_a}$, where $n_1=3$ and $n_\pm=1$, and 
$\tau_1=\frac{\eta}{\lambda}$, 
$\tau_{\pm}=\frac{3 \left[\lambda  \chi _2 c_s^2+\chi _3 \left(\lambda +\chi _2\right)\right]+\chi _1 \left(4 \eta -3 \chi _4\right)\pm\sqrt{\Delta}}{6\lambda\chi_1}$. The existence of real 
roots demands $\lambda,\chi_1>0$, and
\begin{align}
\label{9}
\Delta & =9 \lambda ^2 \chi _2^2 c_s^4+6 \lambda  c_s^2 \left[\chi _1 \left(4 \eta -3 \chi _4\right) \left(2 \lambda +\chi _2\right)+3 \chi _2 \chi _3 \left(\lambda +\chi_2\right)\right] 
\nonumber 
\\ &+\left[\chi _1 \left(4 \eta -3 \chi _4\right)+3 \chi _3 \left(\lambda +\chi_2\right)\right]{}^2\ge0.
\end{align}
In order to fulfill the aforementioned conditions of causality, we need to impose that $0\le\tau_a\le 1$, which gives the following conditions: $\lambda,\chi_1>0$, $\eta\ge0$
\bml
\label{10}
\bea
&&\lambda\ge \eta,\label{10a}\\
&&c_s^2(3\chi_4- 4\eta)\ge 0,\label{10b}\\
&&\lambda\chi_1+c_s^2\lambda\left (\chi_4-\frac{4\eta}{3}\right )\ge c_s^2\lambda \chi_2 + \lambda\chi_3 + \chi_2 \chi_3 - \chi_1 \left (\chi_4-\frac{4}{3} \eta\right) \ge0,\label{10c}\\
&&2\lambda\chi_1 \ge c_s^2\lambda \chi_2 + \lambda\chi_3 + \chi_2 \chi_3 - \chi_1 \left (\chi_4-\frac{4}{3} \eta\right ).\label{10d}
\eea
\eml
Therefore, the Einstein+viscous fluid system in \eqref{4} is causal in the nonlinear regime when $\lambda,\chi_1>0$, $\eta\ge0$, and conditions \eqref{9} and \eqref{10} are satisfied. Note that, in particular, for $c_s=0$ the condition \eqref{10b} is automatically satisfied and does not impose any new constraint on $\chi_4$ and $\eta$. This completes the causality proof (see also Appendix \ref{Appendix_causality} for further mathematical details). The same holds in Minkowski space-time. We note that the fact that $\lambda,\chi_1 >0$ implies that heat flow and non-equilibrium corrections to the energy density must be included for nonlinear causality to hold in a viscous fluid, which explains why Landau-Lifshitz theory \cite{LandauLifshitzFluids} 
(where those terms are omitted) is acausal. 

We conclude this section with the following important remark. 
The following criteria has been used in the literature as 
a test for causality:
$\omega_{sound}(k)$ and $\omega_{shear}(k)$ cannot grow faster than $|k|$
for $|k| \gg 1$ \cite{Pu:2009fj}.
We stress that this simple test  
is restricted only to the \emph{linear} regime and may only \emph{suggest} causality violation. As a matter of fact, there are well-known calculations in causal microscopic theories where $\omega(k) \sim \beta |k|$ with $\beta >1$ for large $|k|$, as found for instance in Ref.\ \cite{Kovtun:2005ev}. 
In contrast with other works that relied on 
tentative linear tests \cite{Kovtun:2019hdm,Van:2007pw,Van:2011yn}, here we 
provide the first full proof of causality, valid even at the nonlinear level, in general first-order theories at zero chemical potential.

%%%%%%%%%%%%%% SECTION  %%%%%%%%%%%%%%%%%%%%%%%%%%%%

\subsection{Linear stability} 

We follow \cite{Hiscock_Lindblom_instability_1985,BemficaDisconziNoronha} and consider small fluctuations around global equilibrium in flat space-time, i.e., $\varepsilon \to \varepsilon + \delta\varepsilon$ and $u^\mu \to u^\mu + \delta u^\mu$ ($u_\mu \delta u^\mu=0$) with $u^\mu = \gamma(1,v^i)$, $\gamma=1/\sqrt{1-v^2}$ ($v^2 = v^i v_i$), and $0 \leq v <1$. After linearizing the fluid equations of motion, we define $\delta\bar{\varepsilon}=\delta\varepsilon/(\varepsilon+P)$ and consider plane wave solutions $\delta \bar{\varepsilon},\delta u^\alpha\to e^{T(\Gamma t+i k^i x_i)}\delta \bar{\varepsilon},\delta u^\alpha$, where $k^\mu=(i\Gamma,k^i)$ (we include $T$ in the exponent to make $k^\mu$ dimensionless). We recall that linear stability demands that the real part $\Re(\Gamma)\leq 0$ for any (constant and uniform) background velocity $v^i$.
For simplicity, we first write the equations in the rest frame where $v=0$. Using $k^2=k^i k_i$
and following \cite{Hiscock_Lindblom_instability_1985}, the 
equations determining the perturbed modes split into two channels:
\bea
\text{Shear channel:}\quad && \bar{\lambda}\Gamma^2+\bar{\eta}k^2+\Gamma=0,\label{15}\\
\text{Sound channel:}\quad && A_0+A_1\Gamma+A_2\Gamma^2+A_3\Gamma^3+A_4\Gamma^4=0,\label{16}
\eea
where $A_0=k^2 c_s^2+\frac{c_s^2}{3} \bar{\lambda}  k^4 \left(3 \bar{\chi}_4-4 \bar{\eta} \right)$, $A_1=\frac{1}{3} k^2 \left[3 c_s^2 \left(\bar{\lambda} +\bar{\chi} _2\right)+4 \bar{\eta} +3 \bar{\chi} _3-3 \bar{\chi} _4\right]$, $A_2=1+k^2\left [\bar{\lambda} \bar{\chi} _3+c_s^2\bar{\chi} _2\bar{\lambda} +\bar{\chi} _2\bar{\chi} _3-\bar{\chi} _1 \left(\bar{\chi} _4-\frac{4 \bar{\eta}}{3}\right)\right ]$, $A_3=\bar{\lambda}+\bar{\chi}_1$, $A_4=\bar{\lambda} \bar{\chi}_1$, and $\bar{\chi}_a=T\chi_a/(\varepsilon+P)$, $\bar{\eta}=T\eta/(\varepsilon+P)$, and $\bar{\lambda}=T\lambda/(\varepsilon+P)$ are dimensionless quantities. The corresponding polynomials when $v^i \neq 0$ can be obtained via a boost, which amounts to changing $\Gamma\to \gamma(\Gamma+i k^iv_i)\quad \text{and}\quad k^2\to -\gamma^2(\Gamma+i k^iv_i)^2+\Gamma^2+k^2$. 

For the shear channel, it is straightforward to prove analytically that condition \eqref{10a}, found to ensure causality, implies stability for any $v^i$. A comparison to similar studies in MIS theory \cite{Romatschke:2017ejr} shows that $\lambda/(\varepsilon+P)$ plays the role of a shear relaxation time. The analysis of the sound channel is more complicated. In the rest frame, real $\Gamma$-roots demand $A_i\ge0$. This is guaranteed by the causality conditions \eqref{10b}, \eqref{10c}, $\lambda$, $\chi_1>0$, and $\eta\ge 0$ together with $c_s^2 \left(\lambda +\chi _2\right)+\frac{4 \eta}{3} + \chi_3-\chi_4\ge0$. Taking $\Gamma=\Gamma_R+i\Gamma_I$ one may use the Routh-Hurwitz criterion \cite{gradshteyn2007} to obtain that $\Gamma_R \leq 0$ imposes the following conditions: Eq.\ \eqref{9} together with
\bml
\label{19}
\bea
&&\zeta+\frac{4 \eta}{3}\ge 0,\label{19a}\\ 
&&3 c_s^2 \{\chi _1 \left[\lambda ^2 \left(4 \eta -3 \chi _4\right)+3 \chi _3 \left(-\lambda ^2+\lambda  \chi _2+\chi
   _2^2\right)\right]+\lambda  [\lambda ^2 \left(4 \eta +3 \chi _3-3 \chi _4\right)+3 \chi _2^2 \chi _3\nonumber\\
&&+\lambda  \chi _2 \left(4 \eta +9 \chi _3-3 \chi_4\right)]+\chi _1^2 \left(4 \eta -3 \chi _4\right) \left(2 \lambda +\chi _2\right)\}-9c_s^4 \lambda ^2\left(\chi _1-\chi _2\right) \left(\lambda +\chi _2\right)\nonumber\\
&&+\left(4 \eta +3 \chi _3-3 \chi _4\right) \left(\chi _1^2 \left(4 \eta
   -3 \chi _4\right)+3 \lambda  \chi _3 \left(\lambda +\chi _2\right)+3 \chi _2 \chi _3 \chi _1\right)\ge 0.
\eea
\eml 
It is worth mentioning that these conditions coincide with the ones obtained in Ref.\ \cite{Kovtun:2019hdm} for the rest frame. Nevertheless, the equal sign in the above inequalities has been included in order to incorporate the also stable situation $\Re(\Gamma)=0$. Since the case where $\eta = 0$ is well-defined, in general \eqref{19a} is satisfied if $\zeta\geq 0$, in accordance with non-negative entropy production. When $v^i \neq 0$,  in the homogeneous $k=0$ case (which corresponds to the lowest order contribution to the dispersion relation $\omega(k)=\omega(0)+\mathcal{O}(k)$ for the sound waves parallel to $v^i$, where $\omega(0)\neq 0$) the stability conditions are: \eqref{9}, \eqref{10b}, \eqref{10c}, and $(\lambda +\chi _1)(1-c_s^2)-\zeta-\frac{4 \eta}{3}\ge0$. Note that in Ref. \cite{Kovtun:2019hdm} the stability conditions for the boosted frame have been verified only for the first and second lowest orders in $k$ for the dispersion relation $\omega(k)$ for the sound waves perpendicular to $v^i$, which does not demand any new condition besides \eqref{19a}. In this sense, the conditions coming from the  homogeneous frame are essential and make a direct link between linear stability and \emph{nonlinear} causality.   
In the non-homogeneous case with $v^i\neq 0$, one is left with a very complex
polynomial that cannot be analyzed analytically. In this case we can still carry out 
the stability analysis numerically, and we did verified stability for several possible
choices of parameters. An extensive numerical study of stability, 
however, is beyond the scope of the present
work and we believe that it is better to investigate stability on a case-by-case basis,
where one already has a pre-determined range of parameter values relevant for specific
applications. 

{\bf Special case where $c_s=0$:} In particular, when $c_s=0$, the condition \eqref{10b} may be dropped and all the above conditions for nonlinear causality and linear stability are satisfied if $\lambda,\chi_1>0$, $\eta\ge 0$, $\lambda\ge \eta$, and
\bml
\bea
\lambda \chi_1\ge \lambda \chi_3+ \chi_2\chi_3+\chi_1\left (\frac{4\eta}{3}-\chi_4\right )\ge 0,\\
\chi_1^2\left (\frac{4\eta}{3}-\chi_4\right )+\lambda\chi_3(\lambda+\chi_2)+\chi_1\chi_2\chi_3\ge0,\\
\lambda+\chi_1\ge\chi_3+\frac{4\eta}{3}-\chi_4\ge 0.
\eea
\eml
%%%%%%%%%%%%%% SECTION  %%%%%%%%%%%%%%%%%%%%%%%%%%%%

\subsection{Local existence and uniqueness} 

We can also establish local existence and uniqueness of solutions to the system of equations
\eqref{4}. The proof relies on techniques
of Leray systems (see \cite{DisconziFollowupBemficaNoronha}). The statement of local existence and uniqueness 
can be summarized as follows. Given sufficiently regular 
initial conditions  for the system of equations
\eqref{4}, there exists a unique solution to \eqref{4}. We refer the reader to 
Appendix \ref{Appendix_causality} for a mathematically rigorous statement and its proof. We remark that while this result is of a
mathematical nature, its importance in physics cannot be underestimated. Not only
are proofs of local existence and uniqueness crucial to provide a solid foundation for the formal
aspects of a theory, but the reliability of numerical simulations might be called
into question absent such proofs \cite{GuermondetalNumerical}.

%%%%%%%%%%%%%%% SECTION %%%%%%%%%%%%%%%%%%%%%%%%%%%%

\section{Conclusions} \label{conclusions}

In this work we derived the most general energy-momentum tensor of a viscous fluid with an arbitrary equation of state, without further conserved currents, that is first-order in the derivatives of the energy density and flow velocity and does not include extended variables such as in Mueller-Israel-Stewart-like theories. We showed that if a choice of hydrodynamic variables distinct from the ones introduced by Eckart and Landau-Lifshitz is adopted, this energy-momentum tensor gives rise to a causal theory. Local existence and uniqueness
of solutions has also been established. These results hold with or without coupling to Einstein's
equations and have been rigorously established. We also showed that linear perturbations
of equilibrium states are stable. A kinetic theory realization of such energy-momentum tensor was also provided. The physical and mathematical properties of the generalization of \eqref{novotensorBDNfodaotche} that includes the effects from a nonzero chemical potential will be the scope of a future work \cite{newBDN} (the general form of the energy-momentum tensor and the conserved current to first-order can already be found in the work of Kovtun \cite{Kovtun:2019hdm}). 
%Furthermore, the cosmological consequences of the general first-order theory of relativistic 
%viscous fluids considered here will be presented elsewhere.

Our results are of relevance for the study of the non-equilibrium dynamics of the quark-gluon plasma formed in heavy-ion collisions. The space-time evolution of this highly dense matter is currently described using MIS theories \cite{Romatschke:2017ejr}, which may be seen as an approximate way to describe the interactions between the hydrodynamic degrees of freedom and the other (faster) degrees of freedom present in the system. After Ref.\ \cite{Heller:2013fn} showed that the gradient expansion can diverge in rapidly expanding systems (see also \cite{Buchel:2016cbj,Denicol:2016bjh,Heller:2016rtz}), attractor dynamics has been proposed \cite{Heller:2015dha} as a way to provide a broader definition of hydrodynamic behavior that can be extended toward the far-from-equilibrium regime \cite{Florkowski:2017olj}. The emergence of a hydrodynamic attractor in the system would then mark the time after which dissipative contributions to the energy-momentum tensor could be reliably described in terms of constitutive relations involving the gradients of the hydrodynamic variables. It is known that MIS theories \cite{Heller:2015dha,Denicol:2017lxn,Denicol:2018pak} and anisotropic hydrodynamics \cite{Strickland:2017kux} display attractor behavior under highly symmetrical flow conditions. Ref.\ \cite{BemficaDisconziNoronha} already showed that the conformal version of the general first-order theory derived here displays a similar attractor behavior. Future work will reveal how the powerful constraints derived here from nonlinear causality, existence, uniqueness, and stability affect the properties of the hydrodynamic attractor of the new theory studied here that contains shear, bulk, and heat flow contributions.   

Our study opens the door for the investigation of several important problems that require
a casual, linearly stable, and local well-posed theory of relativistic viscous fluids, 
such as the study of neutron star mergers, the formation of shocks 
in relativistic viscous fluids, and the generalization, to the viscous context, of known mathematical results valid for perfect fluids. We hope to be able to
address some of these questions in the near future.

%%%%%%%%%%%%%%%%%%%%%%%%%%%%%%%%%%%%%%%%%%%%

\section*{Acknowledgements} 

We thank P.~Kovtun for showing the results of Ref.\ \cite{Kovtun:2019hdm} to us before submission. FSB is partially supported by a Discovery Grant administered by Vanderbilt University. Part of this work was done while FSB was visiting Vanderbilt University. MMD is partially supported by a Sloan Research Fellowship provided by the Alfred P. Sloan foundation, NSF grant DMS-1812826, 
a Discovery Grant administered by Vanderbilt University, and a Dean's Faculty Fellowship. 
JN was partially supported by 
CNPq grant 306795/2017-5 and FAPESP grant 2017/05685-2. 

\appendix

\section{Kinetic theory derivation}\label{Appendix_kinetic}

Following \cite{Denicol:2012cn}, we consider the Boltzmann equation for a dilute relativistic gas of (single species) particles with constant mass $\mathcal{M}$ (in flat space-time)\footnote{The coupling with gravity is straightforward, see \cite{kremer}.}
\be
\label{2kinetic}
k^\mu\nabla_\mu f_k(x)=\mathcal{C}[f_k]\,,
\ee
where $\mathcal{C}[f_k]$ is the collision kernel and $f_k(x)=f(k,x)$ is the distribution function that depends on the space-time coordinates $x^\mu$ and on the on-shell momenta $k^\mu=(k^0,k^i)$, with $k^0=\sqrt{k^i k_i+\mathcal{M}^2}$. From $f_k$ we may define quantities such as the energy-momentum tensor 
\be
\label{EM-tensor}
T^{\mu\nu}(x)=\Bracket{k^\mu k^\nu},
\ee 
where $\Bracket{h_k}$ stands for \[\Bracket{h_k}=\int_k f_k h_k\] for any function $h_k$. We also define $\int_k=\int\frac{d^3k}{(2\pi)^3k^0}$, with $\frac{d^3k}{(2\pi)^3k^0}$ being the Lorentz invariant measure. We focus here on the derivation of $T^{\mu\nu}$.  

The collision kernel is given by
\be
\label{3kinetic}
\mathcal{C}[f_k]=\frac{1}{2}\int_{k^\prime\,p\,p^\prime}W(kk^\prime|pp^\prime)(f_pf_{p^\prime}-f_kf_{k^\prime})\,,
\ee
where $W(kk^\prime|pp^\prime)$ is the Lorentz invariant transition rate for (elastic) 2 to 2 collisions\footnote{Though the exact form of $W(kk^\prime|pp^\prime)$ is not important in the following, we assume that the standard properties needed for the H-theorem to hold \cite{kremer} are valid.}. For simplicity, in this work we neglect effects from quantum statistics and consider classical statistics, as this does not affect the important steps needed in the derivation of the energy-momentum tensor. The collision kernel obeys the relations \[\int_k \mathcal{C}[f_k]=\int_k k^\mu \mathcal{C}[f_k]=0,\] which define the conservation laws. 

We note that any distribution function of the form
\be
e^{k^\mu \xi_\mu/\vartheta + \varphi}
\ee
with $\xi_\mu, \vartheta,\varphi$ being at this point arbitrary (normalized) time-like vector and scalar fields, respectively, is a zero of the collision kernel, i.e., $\mathcal{C}[e^{k^\mu \xi_\mu/\vartheta + \varphi}] = 0$. However, such a distribution is only a solution of the Boltzmann equation if the left-hand side is also zero, i.e., if the fields obey
\be
k^\nu \nabla_\nu \left(e^{k^\mu \xi_\mu/\vartheta + \varphi}\right)=0,
\ee
which implies that 
\be
\nabla_\mu \varphi=0 \qquad \textrm{and} \qquad \nabla_\mu \left(\xi_\nu/\vartheta\right) +\nabla_\nu \left(\xi_\mu/\vartheta\right) =0
\label{eqcond1}
\ee
so $\xi_\nu/\vartheta$ is a Killing vector field  \cite{kremer}. The fields $\{\xi_\nu, \vartheta,\varphi\}$ may then be identified with the standard hydrodynamic variables $\{u_\nu, T,\mu\}$ of ideal hydrodynamics and \eqref{eqcond1} can be written as
\bea
&&\sigma_{\mu\nu} = 0 \\
&&\nabla_\mu u^\mu=0 \qquad \textrm{and} \qquad u^\nu \nabla_\nu\mu=0 \qquad \textrm{and} \qquad  u^\nu \nabla_\nu T=0 \\
&&u^\nu \nabla_\nu u_\mu + \frac{\nabla_\mu T}{T}=0\\
&&\Delta^{\nu}_\beta\nabla_\nu(\mu/T)=0,
\eea
which are the standard conditions that define thermodynamic equilibrium. It should be clear  from the derivation above that $\{u_\nu, T,\mu\}$ are not uniquely defined. In fact, it is more adequate to say that there are is an infinite number of equilibrium states that satisfy the Boltzmann equation. 

From now on, we set the chemical potential to zero $(\mu=0)$ and denote this class of equilibrium distributions by
\be
\label{1kinetic}
f^{eq}_k(x)=e^{-E_k/T}
\ee
where $E_k=-u^\alpha k_\alpha$. Besides the flow velocity, the equilibrium part of the energy-momentum tensor involves  
\[\varepsilon=\Bracket{E_k^2}_{eq}\quad\text{and}\quad P=\frac{1}{3}\Delta_{\mu\nu}\Bracket{k^\mu k^\nu}_{eq},\] where $\Bracket{h_k}_{eq}$ denotes \[\Bracket{h_k}_{eq}=\int_k f^{eq}_k h_k.\]
It is convenient to also define the variation \[\Bracket{h_k}_\delta=\Bracket{h_k}-\Bracket{h_k}_{eq}\] and perform the decompositions $k^\mu=E_k u^\mu+\kappa^\mu$ and 
\be
\label{4kinetic}
k^\mu k^\nu=E_k^2u^\mu u^\nu+E_k u^\mu \kappa^\nu+E_k u^\nu \kappa^\mu+k^{\langle\mu}k^{\nu\rangle}+\frac{\Delta^{\mu\nu}}{3}\kappa^2\,,
\ee
where $\kappa^\mu=\Delta^{\mu\nu}k_\nu$, $\kappa^2=\kappa^\alpha\kappa_\alpha=E_k^2-\mathcal{M}^2$, and $k^{\langle\mu}k^{\nu\rangle}=\Delta^{\mu\nu}_{\alpha\beta}k^\alpha k^\beta$, with $\Delta^{\mu\nu}_{\alpha\beta}=(1/2)[\Delta^\mu_\alpha\Delta^\nu_\beta+\Delta^\nu_\alpha\Delta^\mu_\beta-(2/3)\Delta^{\mu\nu}\Delta_{\alpha\beta}]$. Then, the most general form for $T^{\mu\nu}$ that includes out-of-equilibrium contributions is
\bea
T^{\mu\nu}&=&\left (\varepsilon+\Bracket{E_k^2}_\delta\right )u^\mu u^\nu+\left (P+\frac{\Bracket{\kappa^2}_\delta}{3}\right )\Delta^{\mu\nu}+\Bracket{\kappa^{\langle\mu}\kappa^{\nu\rangle}}_\delta+u^\mu \Bracket{E_k\kappa^\nu}_\delta+u^\nu\Bracket{E_k\kappa^\mu}_\delta,
\label{5kinetic}
\eea
where $\Bracket{E_k^r\kappa^\mu}_{eq}=0$ and $\Bracket{\kappa^{\langle\mu}\kappa^{\nu\rangle}}_{eq}=0$ by symmetry.

We follow the approximations discussed in \cite{BemficaDisconziNoronha} and consider perturbations around local equilibrium by setting $f_k\approx f^{eq}_k+\delta f_k$, where $\delta f_k=f^{eq}_k\phi_k(x)$. Then, up to first-order in $\delta f_k$, Eq.\ \eqref{2kinetic} may be written as
\be
\label{6kinetic}
k^\mu\nabla_\mu f^{eq}_k+k^\mu\nabla_\mu \left (f^{eq}_k\phi_k\right )=f^{eq}_k\mathcal{L}[\phi_k],
\ee  
where
\be
\label{7kinetic}
\mathcal{L}[\phi_k]=\frac{1}{2}\int_{pp^\prime k^\prime}W(kk^\prime|pp^\prime)f^{eq}_{k^\prime}\left (\phi_p+\phi_{p^\prime}-\phi_k-\phi_{k^\prime}\right )
\ee
is an operator with kernel spanned by the set $\{1,E_k,\kappa^\mu\}$ that obeys $\Bracket{h_k\mathcal{L}[z_k]}_{eq}=\Bracket{z_k\mathcal{L}[h_k]}_{eq}$ and  $\Bracket{h_k\mathcal{L}[h_k]}_{eq}<0$. We assume that $\phi_k(x)$ is first-order in the derivatives of $T$ and $u^\mu$. Keeping only terms that are first-order in derivatives, the solution of Eq.\ \eqref{6kinetic} can be obtained from the moments \cite{BemficaDisconziNoronha} 
\be
\label{8kinetic}
\int_k k^{j_1}\cdots k^{j_n}\left \{k^\mu\nabla_\mu f^{eq}_k-f^{eq}_k\mathcal{L}[\phi_k]\right \}=0,
\ee
where $j=0,1,\cdots$. In particular, for $j=0,1$ one obtains the conservation laws. As for $j=2$, by means of \eqref{4kinetic} and
\bea
\label{9kinetic}
u^\mu\nabla_\mu f^{eq}_k&=&f^{eq}_k\Bigg \{\frac{k^{\langle\mu}k^{\nu\rangle}\sigma_{\mu\nu}}{T}+\kappa^\mu\left [\frac{E_k}{T}\left (\frac{\nabla^\perp_\mu T}{T}+u^\alpha\nabla_\alpha u_\mu\right )\right ]+\frac{E_k^2 u^\alpha\nabla_\alpha T}{T^2}\nonumber\\
&&+\frac{\kappa^2 \nabla_\alpha u^\alpha}{3T}\Bigg \}
\eea
we obtain the equations
\bml
\label{10kinetic}
\bea
&&I_4A=\frac{1}{T^5}\Bracket{E_k^2\mathcal{L}[ \phi_k]}_{eq},\\
&&\frac{L_{2,2}}{3} q^\mu
=\frac{1}{T^5}\Bracket{E_k\kappa^\mu\mathcal{L}[\phi_k]}_{eq},\\
&&\frac{2L_{0,4}}{15}\Delta^{\mu\nu\alpha\beta}\sigma_{\alpha\beta}=\frac{1}{T^5}\Bracket{k^{\langle\mu}k^{\nu\rangle}\mathcal{L}[\phi_k]}_{eq}
\eea
\eml
where 
\bml
\label{13kinetic}
\bea
A&=&\frac{u^\alpha\nabla_\alpha T}{T}+l\, \nabla_\alpha u^\alpha,\\
q^\mu&=&\frac{\nabla_\perp^\mu T}{T}+ u^\alpha\nabla_\alpha u^\mu,
\eea
\eml
with
\bml
\label{11kinetic}
\bea
l&=&\frac{L_{2,2}}{3I_4}=\frac{1}{3}-\frac{\mathcal{M}^2}{T^ 2}\frac{I_2}{3I_4}.\label{11a}
\eea
\eml
The dimensionless integrals above are defined as (using the fact that $\kappa^2=\Delta^{\alpha\beta}k_\alpha k_\beta\ge0$)
\[I_n=\frac{\Bracket{E_k^n}_{eq}}{T^{n+2}}>0\quad\text{and}\quad L_{n,m}=\frac{\Bracket{E_k^n\kappa^m}_{eq}}{T^{n+m+2}}>0.\] The kernel of the operator $\mathcal{L}$ is a subspace of dimension 5, which implies that $\phi_k$ is not uniquely obtained from \eqref{10kinetic}. Actually, one may write $\phi_k=\phi_k^{(p)}+\phi_k^{{h}}$, where the homogeneous part $\phi_k^{(h)}\in\ker(\mathcal{L})$ ($\mathcal{L}[\phi_k]=\mathcal{L}[\phi_k^{(p)}]$), with the particular solution $\phi_k^{(p)}$ being completely determined by \eqref{10kinetic}. Thus, the most general $\phi_k$ that satisfies \eqref{10kinetic} is
\be
\label{general_phi}
\phi_k=\phi_A\frac{k^{\langle\mu}k^{\nu\rangle}\sigma_{\mu\nu}}{T^3}+\phi_B\frac{E_k^2 A}{T^3}+\phi_C\frac{E_k \kappa^\mu }{T^3}q_\mu-\left (\phi_B\frac{B}{T}+\phi_B\frac{C}{T}\frac{E_k}{T} +\phi_C \frac{\kappa^\mu}{T^2}D^\perp_\mu\right ),
\ee
where in parentheses we wrote the homogeneous terms as combinations of $1,E_k$, and $\kappa^\mu$, while the particular solution is uniquely determined by $A$ and $q^\mu$ from \eqref{10kinetic}. The most general form of the homogeneous terms must be combinations of quantities that vanish in equilibrium, i.e, $\nabla_\alpha u^\alpha$, $u^\alpha\nabla_\alpha T$, and $\nabla^\mu_\perp T/T+u^\alpha\nabla_\alpha u^\mu$, and thus  
\bml
\label{12kinetic}
\bea
B&=&b_1\frac{u^\alpha\nabla_\alpha T}{T}+b_2\, \nabla_\alpha u^\alpha,\\
C&=&c_1\frac{u^\alpha\nabla_\alpha T}{T}+c_2\, \nabla_\alpha u^\alpha,\\
D^\mu_\perp&=& d\left (\frac{\nabla_\perp^\mu T}{T}+ u^\alpha\nabla_\alpha u^\mu\right ),
\eea
\eml
where the dimensionless coefficients $b_i$, $c_i$, and $d$ define the terms that enter in the first-order theory (and also the sign of its coefficients). This is how our choice of \emph{hydrodynamic frame} appears in the context of kinetic theory, which nicely provides a microscopic realization of the ideas presented by Kovtun in \cite{Kovtun:2019hdm}. The quantities $\phi_A$, $\phi_B$, and $\phi_C$ contain the independent information regarding transport  and they can be obtained by using \eqref{general_phi} into \eqref{10kinetic} and then solving the following equations:
\bml
\label{14kinetic}
\bea
&&\frac{2L_{0,4}}{15}\Delta^{\mu\nu\alpha\beta}\sigma_{\alpha\beta}=\frac{\phi_A}{T^8}\Bracket{k^{\langle\mu}k^{\nu\rangle}\mathcal{L}[k^{\langle\alpha}k^{\beta\rangle}]}_{eq}\sigma_{\alpha\beta},\\
&&I_4=\frac{\phi_B}{T^8}\Bracket{E_k^2\mathcal{L}[ E_k^2]}_{eq},\\
&&\frac{L_{2,2}}{3}q^\mu
=\frac{\phi_C}{T^8}\Bracket{E_k\kappa^\mu\mathcal{L}[E_k\kappa^\nu]}_{eq}q_\nu.
\eea
\eml
Exact expressions for the transport coefficients depend on $\phi_A$, $\phi_B$, and $\phi_C$, which can be found once the microscopic details involving the particle scattering are given. However, in this work we will not focus on such calculations. Rather, our goal here is only to determine their general properties. First, we remark that \eqref{14kinetic} implies that $\phi_A,\phi_B,\phi_C<0$ since $\Bracket{h_k\mathcal{L}[h_k]}_{eq}<0$. Then, given that $\Bracket{h_k}_\delta=\Bracket{h_k \phi_k}_{eq}$, one obtains
\bml
\label{15kinetic}
\bea
\MA_1&=&\Bracket{E_k^2\phi_k}_{eq}=-T^3\phi_B\bigg [\left (b_1 I_2+c_1 I_3-I_4\right )\frac{u^\alpha \nabla_\alpha T}{T}+\left (b_2 I_2+c_2 I_3-lI_4\right )\nabla_\alpha u^\alpha\bigg ],\\
\MA_2&=&\frac{1}{3}\Bracket{\kappa^2\phi_k}_{eq}=-\frac{T^3\phi_B}{3}\bigg [\left (b_1 L_{0,2}+c_1 L_{1,2}-L_{2,2}\right )\frac{u^\alpha \nabla_\alpha T}{T}+\big(b_2 L_{0,2}+c_2 L_{1,2}\nonumber\\
&&-l L_{2,2} \big)\nabla_\alpha u^\alpha\bigg ],\\
\mathcal{Q}^\mu&=&\Bracket{E_k\kappa^\mu\phi_k}_{eq}=-\frac{T^3\phi_C}{3}\left ( L_{1,2}\,d-L_{2,2}\right )\bigg (\frac{\nabla_\perp^\mu T}{T}+u^\alpha \nabla_\alpha u^\mu\bigg ),\\
\eta\sigma^{\mu\nu}&=&-\frac{1}{2}\Bracket{k^{\langle\mu}k^{\nu\rangle}\phi_k}_{eq}=-T^3\phi_A\frac{L_{0,4}}{15}\sigma^{\mu\nu}.
\eea
\eml
One can obtain immediately that \[\eta=-\phi_A\frac{L_{0,4}}{15T^3}>0.\] 

Now, it is easy to see that the energy-momentum tensor discussed in this paper 
\bea
T^{\mu\nu}&=&(\varepsilon+\MA_1)u^\mu u^\nu+\left (P+\MA_2\right )\Delta^{\mu\nu}-2\eta\sigma^{\mu\nu}+u^\mu\mathcal{Q}^\nu+u^\nu\mathcal{Q}^\mu,\label{tensores}
\eea  
where
\bml
\label{Def}
\bea
\MA_1&=&\frac{\chi_1}{c_s^2} \frac{u^\alpha\nabla_\alpha T}{T}+\chi_2 \nabla_\alpha u^\alpha,\label{def1}\\
\MA_2&=&\frac{\chi_3}{c_s^2}\frac{u^\alpha\nabla_\alpha T}{T}+\chi_4 \nabla_\alpha u^\alpha,\label{def1-2}\\
\mathcal{Q}_\mu&=&\lambda\left (\frac{\nabla^\perp_\mu T}{T}+u^\alpha\nabla_\alpha u_\mu\right ),\label{def2}
\eea
\eml
with $\nabla^\perp_\mu=\Delta_\mu^\nu\nabla_\nu$ and $dT/T=dP/(\varepsilon+P)=c_s^2 d\varepsilon/(\varepsilon+P)$, is obtained when one sets  
\bml
\label{16kinetic}
\bea
\frac{\chi_1}{c_s^2}&=&-m_1T^3\phi_B,\\
\chi_2&=& -m_2T^3\phi_B ,\\
\frac{\chi_3}{c_s^2}&=&-m_3\frac{T^3\phi_B}{3} ,\\
\chi_4&=&-m_4\frac{T^3\phi_B}{ 3},\\ 
\lambda&=&-r\frac{T^3\phi_C}{3},
\eea
\eml
where $m_i$ and $r$ are chosen positive such that
\bml
\label{17kinetic}
\bea
&&b_1 I_2+c_1 I_3-I_4=m_1>0,\\
&&b_1 L_{0,2}+c_1 L_{1,2}-L_{2,2}=m_3>0\\
&&b_2 I_2+c_2 I_3-lI_4=m_2>0,\\
&&b_2L_{0,2}+c_2 L_{1,2}-l L_{2,2}=m_4>0,\\
&& L_{1,2}\,d-L_{2,2}=r>0.
\eea
\eml
Eqs.\ \eqref{17kinetic} fix all the 5 parameters $b_i$, $c_i$, and $d$ in such a way that $\chi_a$'s and $\lambda$ are positive. In the special case where $\mathcal{M}=0$ we must have from \eqref{17kinetic} that $m_1=m_3$ and $m_2=m_4$ since $L_{m,n}=I_{m+n}$ and $l=c_s^2=1/3$ [see \eqref{11a}]. Also, in this case $m_1=3m_2$ so that  $\chi_4=\chi_3=\chi_2/3=\chi_1/3$, which reduces to the case considered in \cite{BemficaDisconziNoronha}.

\newpage

\section{Formal proof of causality, local existence, and uniqueness}\label{Appendix_causality}

Here we provide the formal proofs of the statements of local existence, uniqueness, and
causality for Einstein's equations coupled to 
\eqref{novotensorBDNfodaotche}. We use the standard terminology of general 
relativity (see, e.g., \cite{HawkingEllisBook}). An initial data set
for  Einstein's equations coupled to 
\eqref{novotensorBDNfodaotche} consists of a three-dimensional manifold
$\Sigma$, a Riemannian metric $\mathring{g}$ and a symmetric two tensor
$\kappa$ on $\Sigma$, two vector fields $\mathring{u}$ and $\mathring{U}$ on $\Sigma$,
and two scalar functions $\mathring{\varepsilon}$ and $\mathring{\mathcal{E}}$ on 
$\Sigma$, such that the Einstein constraint equations are satisfied.
The fields $\mathring{u}$ and $\mathring{\varepsilon}$ correspond 
to $\left.u^i\right|_{t=0}$ and $\left. \varepsilon \right|_{t=0}$, respectively,
whereas 
$\mathring{U}$ and $\mathring{\mathcal{E}}$ correspond 
to $\left. \partial_t u^i\right|_{t=0}$ and $\left. \partial_t \varepsilon \right|_{t=0}$, respectively. We note that only initial data for the projection of $u$  
onto the tangent bundle of $\Sigma$ is given initially in view of the normalization condition
$u^\alpha u_\alpha = -1$. Similarly for transversal (to $\Sigma$) derivatives of $u$. 
In Theorem I below, $G^{s}$ is the Gevrey space (see, e.g., \cite{RodinoGevreyBook}).
For the proof of Theorem I, we will use techniques of Leray-Ohya systems
developed in 
\cite[\S 6, sec. 27]{LerayOhyaNonlinear} and \cite{CB_diagonal}. A statement of the result as needed here appears in \cite[Appendix A]{DisconziFollowupBemficaNoronha}
(see also \cite[p. 624]{ChoquetBruhatGRBook} for a simplified statement).

\vskip 0.1cm
\textbf{Theorem I.} 
Consider the energy-momentum tensor \eqref{novotensorBDNfodaotche} and assume that
 $\lambda$, $\chi_a$, $a=1,\dots, 4$, $\eta$,
and $P$ are given real valued functions with domain $(0,\infty)$,
where we recall that in \eqref{novotensorBDNfodaotche} 
these quantities are functions of $\varepsilon$, i.e., 
$\lambda = \lambda(\varepsilon)$, 
$\chi_a = \chi_a(\varepsilon)$, $\eta = \eta(\varepsilon)$, and $P = P(\varepsilon)$.
Suppose that $\lambda$, $\chi_a$, $\eta$, and $P$ are $G^{(s)}$ regular.
Let $\mathcal{I} = (\Sigma, \mathring{\varepsilon}, \mathring{\mathcal{E}},
\mathring{u}, \mathring{U})$ be an initial data set for Einstein's equations
coupled to \eqref{novotensorBDNfodaotche}. 
Assume that the initial data belongs to
$G^{(s)}(\Sigma)$. Suppose that $\Sigma$ is compact and that 
$\mathring{\varepsilon}>0$. 
Suppose that $P^\prime \geq 0$, that 
$\lambda >0$, $\chi_1>0$, $\eta \geq 0$, and that conditions \eqref{9} and \eqref{10} 
hold. Finally, assume that $1 < s < 20/19$.
Then, there exist a four-dimensional Lorentzian manifold $(M,g)$,
a vector field $u$ and a real valued function $\varepsilon$, both defined on $M$, such that:
\vskip 0.1cm
\noindent (1) Einstein's equations coupled to \eqref{novotensorBDNfodaotche}
hold in $M$.
\vskip 0.1cm
\noindent (2) There exists an isometric embedding $i: (\Sigma,\mathring{g}) \rightarrow 
(M,g)$ with second fundamental form $\kappa$.
\vskip 0.1cm
\noindent (3) Identifying $\Sigma$ with its image $i(\Sigma)$ in $M$, we have
$\left. \varepsilon \right|_{\Sigma} = \mathring{\varepsilon}$ and 
$\mathsf{\Pi}_\Sigma(u) = \mathring{u}$, where $\mathsf{\Pi}_\Sigma: \left.TM\right|_{\Sigma}
 \rightarrow T\Sigma$
is the canonical projection from the tangent bundle of $M$ restricted to $\Sigma$ to the tangent bundle 
of $\Sigma$. Furthermore, if $\{ x^\alpha \}_{\alpha=0}^3$ is a system of coordinates
near $\Sigma$ such that $\{x^i\}_{i=1}^3$ are coordinates on $\Sigma$, then
$\left. \partial_0 \varepsilon\right|_{\Sigma} = \mathring{\mathcal{E}}$ and
$\left. \partial_0 u^i \right|_{\Sigma} = \mathring{U}$.
\vskip 0.1cm
\noindent (4) $(M,g)$ is globally hyperbolic with Cauchy surface $i(\Sigma)$.
\vskip 0.1cm
\noindent (5) $(M,g)$ is causal, in the following sense: for any $x$ in the future\footnote{The
future of a set in $M$ is well-defined because $(M,g)$ is globally hyperbolic.} 
of $i(M)$, $(g(x), u(x), \varepsilon(x))$ depends only on $\left.\mathcal{I}\right|_{i(\Sigma)\cap J^-(x)}$, where $J^-(x)$ is the causal past of $x$ (with respect to the metric $g$).
\vskip 0.1cm
\noindent (6) $(M,g)$ is unique up to actions of diffeomorphisms of $M$.
\vskip 0.2cm
\noindent \emph{Proof:} 
We first note that causality, item (5), has already been proved in Section \ref{section_causality}. For, assume that a globally hyperbolic solution exists. Then, the corresponding characteristic manifolds of the Einstein
equations coupled to \eqref{novotensorBDNfodaotche} have been computed in Section
\ref{section_causality}
for Einstein's equations written in wave coordinates. The invariance of the characteristics
\cite[Chapter V]{Courant_and_Hilbert_book_2} assures that causality holds independently 
of the system of coordinates we choose.

In order to establish existence, we embed $\Sigma$ into $\mathbb{R} \times \Sigma$ and consider a coordinate
system $\{x^\alpha\}_{\alpha=0}^3$ in a neighborhood of a point $p \in \Sigma$. Without loss
of generality we can assume that $\{ x^i \}_{i=1}^3$ are coordinates on $\Sigma$ and that
$\mathring{g}(p)$ is the Euclidean metric when expressed in these coordinates. 
We consider
Einstein's equations written in wave gauge, in which case the equations of motion
can be written as in \eqref{4}. As usual in problems for Einstein's equations
in wave gauge, we take as initial conditions for the components of the metric the following:
\begin{gather}
g_{ij}(0,\cdot) = \mathring{g}_{ij}, g_{00}(0,\cdot) = -1,
g_{0i}(0,\cdot) = 0, \partial_0 g_{ij} = \kappa_{ij},
\nonumber
\end{gather}
with $\partial_0 g_{\alpha 0}(0,\cdot)$ chosen such that $\{x^\alpha\}_{\alpha}^3$ are wave
coordinates at $x^0 = 0$. For the fluid variables, we take
\begin{gather}
\varepsilon(0,\cdot) = \mathring{\varepsilon}, \partial_0 \varepsilon(0,\cdot) = 
\mathring{\mathcal{E}},
u^i(0,\cdot) = \mathring{u}^i, \partial_0 u^i(0,\cdot) = \mathring{U}^i,
\nonumber
\end{gather}
with the initial conditions $u^0(0,\cdot)$ and $\partial_0 u^0(0,\cdot)$ determined
from the normalization condition $u^\alpha u_\alpha = -1$.

We group that unknowns $\varepsilon$, $u^\alpha$, and $g_{\mu\nu}$ in the
15-component vector $V=(\varepsilon,u^\alpha,g_{\mu\nu})$. 
To each component $V^I$ we associate an index $m_I$, $I=1,\dots,15$, and to each one of the 15 equations in \eqref{4} we associate an index $n_J$, in such a way that equations
\eqref{4} can be written as
\begin{gather}
 h^J_I(\partial^{m_K-n_J-1}V^K,\partial^{m_I-n_J})V^I+b^J(\partial^{m_K-n_J-1}V^K)=0,
\label{system_general}
\end{gather}
where $I,J=1,\dots,15$, $ h^J_I(\partial^{m_K-n_J-1}V^K,\partial^{m_I-n_J})$
is a homogeneous differential operator of order $m_I - n_J$ (which could possibly be zero)
whose coefficients depend on at most $m_K - n_J - 1$ derivatives of $V^K$, 
$K=1,\dots, 15$, and there is a sum over $I$ in  
$h^J_I(\cdot)V^I$. The terms $b^J(\partial^{m_K-n_J-1}V^K)$ also depend on at most
$m_K - n_J - 1$ derivatives of $V^K$, $K=1,\dots, 15$. The indices $m_I$ and $n_J$ are
defined up to an overall additive constant, but the simplest choice to have equations
\eqref{4} written as \eqref{system_general} is $m_I = 2$, $n_J = 0$, for all 
$I,J=1,\dots,15$.

The characteristic determinant of \eqref{system_general} was computed in
Section \ref{section_causality} and gives
\begin{gather}
\det H(V,\xi) =  \frac{\lambda^4\chi_1}{(\varepsilon+P)}(g^{\mu \nu} \xi_\mu \xi_\nu)^{10}
\prod_{a=1,\pm}\left [(u^ \alpha\xi_\alpha)^2-\tau_a \Delta^{\alpha\beta}\xi_\alpha\xi_\beta\right ]^{n_a},
\nonumber
\end{gather}
where $H = ( h^J_I(\partial^{m_K-n_J-1}V^K,\xi))$ is the characteristic matrix of the system
and the other quantities are as in Section \ref{section_causality}. 
Under our assumptions, the polynomials
$g^{\mu \nu} \xi_\mu \xi_\nu$ 
are hyperbolic polynomials when $V$  takes the initial data. 
Also, when $V$ takes the initial data,
the polynomials 
$(u^ \alpha\xi_\alpha)^2-\tau_a \Delta^{\alpha\beta}\xi_\alpha\xi_\beta$, $a=1, \pm$,
are hyperbolic polynomials for $\tau_a >0$ and products of two hyperbolic
polynomials for $\tau_a=0$. 
Since the roots of a polynomial
are continuous functions of the polynomial coefficients, we conclude that $\det H(V,\xi)$
is a product of at most 20 hyperbolic polynomials for any $V$ 
sufficiently close to the initial data. Moreover, the intersection of the 
characteristic cones defined by these polynomials has non-empty interior. 

Therefore, we have verified the hypotheses of \cite[Theorem A.23]{DisconziFollowupBemficaNoronha} and
we conclude that equations \eqref{4} admit a solution in a neighborhood of $p$. Recall that
a solution to Einstein's equations in wave coordinates gives rise to a solution 
to the full Einstein equations (i.e., Einstein's equations in arbitrary coordinates) if and 
only if the constraint equations are satisfied, which is the case by assumption. Thus,
we have obtained a solution to Einstein's equations coupled to \eqref{novotensorBDNfodaotche}
in a neighborhood of $p$. A standard gluing argument that relies on the causality
of solutions already established (see, e.g., \cite[Chapter 10]{WaldBookGR1984}
or \cite{DisconziFollowupBemficaNoronha}) gives a solution defined in a neighborhood of $\Sigma$.
We have therefore obtained a space-time where statements (1)-(5) hold (we notice that statements
(2)-(4) are immediate consequences of the above constructions). Finally, statement (6)
is obtained by considering the maximal globally hyperbolic development of the initial data
\cite{ChoquetGerochMaximalDevelopment}. \hfill\qed

\vskip 0.2cm
We note that some of the assumptions of Theorem I can be relaxed, but we have not done
so for simplicity. For example, the compactness of $\Sigma$ can be dropped provided
that suitable asymptotic conditions on the fields are given. The Gevrey regularity $1 < s <20/19$
can also be improved. For example, if we are given two hyperbolic polynomials
of the form $p_1(\xi) = (u^ \alpha\xi_\alpha)^2-c_1 \Delta^{\alpha\beta}\xi_\alpha\xi_\beta$
and $p_2(\xi) = (u^ \alpha\xi_\alpha)^2-c_2 \Delta^{\alpha\beta}\xi_\alpha\xi_\beta$ with 
$0< c_1 <c_2 \leq 1$, then the product $p_1(\xi) p_2(\xi)$ is a (degree four) hyperbolic polynomial.
Thus, considering products, we can write $\det H(V,\xi)$
as a product of fewer than 20 polynomials, leading to a better Gevrey regularity
(the range of values of $s$ allowable is determined by $Q/(Q-1)$ when 
$\det H(V,\xi)$ is written as a product of $Q$ hyperbolic polynomials, see
\cite{DisconziFollowupBemficaNoronha}).

Typically for problems in relativity, one wants to establish local existence and uniqueness
under more general regularity assumptions on the initial data than Gevrey regularity.
A common goal is to have a result valid for initial data belonging to Sobolev spaces
\cite{AnileBook}. In this regard, we announce here the following result, which will
be established in the forthcoming paper  \cite{DisconziBDNSobolev}:

\vskip 0.1cm
\textbf{Theorem II.} In Theorem I, assume further that $P^\prime > 0$, 
and suppose that the data belongs to the Sobolev space $H^s$ for sufficiently large $s$.
Then, the same conclusions of Theorem I hold.

The above arguments also show that the fluid equations are locally well-posed in a fixed
background (i.e., without considering coupling to Einstein's equations)

\bibliography{References.bib}

\end{document}